# The Short Anthropological Guide to the Study of Ethical AI

Alexandrine Royer
10th October, 2020

# Table of Contents

## Introduction

In the 1950s, computer scientist Alan Turing designed a test; a machine would be considered 'intelligent' if a human interacting with it could not tell whether it was a person or a machine. It was the first step in the development of what would become the field of Artificial Intelligence (AI), a term first coined by John McCarthy at the seminal Dartmouth summer research project in 1956.[1] In the short span of seventy years, the production of intelligent machines has evolved beyond the scope of human imagination. No longer limited to sci-fi aficionados and the scientific community, artificial intelligence has become ubiquitous in each of our lives. We interact with AI, whether knowingly or unknowingly, daily when using our phones, digital assistants, applying for loans, undergoing medical treatment, or just browsing the web. Companies across the board are scrambling to adopt AI and machine learning technology. Opinions, hopes, and fears ranging from utopia to catastrophe accompany this growing proximity with artificial intelligence systems - Stephen Hawkings'[2] infamously prophesied that AI could spell the end of humanity.

The development of technology has brought on a series of significant advances, such as improved medical imaging, new video communication technology, 3-D printed affordable homes, drones for service deliveries in conflict areas, etc. AI has proven it can produce immense social good[3]. However, every new technology comes with considerable caveats, which we tend to observe once set in motion. The rapid expansion of consumer Internet over the past two decades has led to the explosion of algorithmic decision-making and predictions on individual consumers and behaviour. Before we could even agree to the collection of our data, private corporations, banks, and the public sector used it to make crucial decisions on our lives. Over the years, data scientists and social scientists have started to signal incidents where algorithms violate fundamental social norms and values. Algorithms trampled on notions of privacy, fairness, equality, and were revealed to be prone to manipulations by its users. These problems with algorithms have led researchers Michael Kerns and Aaron Roth to state that "it is less a concern about algorithms becoming more powerful than humans, and more about them altering what it means to be human in the first place."[4]

Over the next few years, society as a whole will need to address what core values it wishes to protect when dealing with technology. Anthropology, a field dedicated to the very notion of what it means to be human, can provide some interesting insights into how to cope and tackle these changes in our Western society and other areas of the world. It can be challenging for social science practitioners to grasp and keep up with the pace of technological innovation, with many being unfamiliar with the jargon of AI. This short guide serves as both an introduction to AI ethics and social science and anthropological perspectives on the development of AI. It intends to provide those unfamiliar with the field with an insight into the societal impact of AI systems and how, in turn, these systems can lead us to rethink how our world operates.

Before delving into anthropology's contributions to AI, a brief overview of the ethical issues in technology will help situate some of the critical failures of algorithmic design and their integration into high-stakes decision-making areas. Exploring the limitations of ethically fine-tuned, or better-behaved, algorithms in the areas of privacy, fairness, and user model manipulation elucidates how ethical AI requires input from the social sciences. The current controversies in which technology giants are enmeshed show that society cannot entrust Silicon Valley entirely to pave the way to produce ethical AI. Therefore, anthropological studies can help determine new avenues and perspectives on how to expand the development of ethical artificial intelligence and machine learning systems. Ethnographic observations have already been used to understand the social contexts in which these systems are designed and deployed. By looking beyond the algorithm and turning to the humans behind it, we can start to critically examine the broader social, economic and political forces at play in the rapid rise of AI and ensure that no population nor individuals are left to bear the negative consequences of technological innovation.

## Brief Overview of the Ethical Issues in Tech

In the past few years, there has been an explosion of ethical concerns raised by technology and its harm to specific groups of people. Researchers have pointed to repeated algorithmic bias cases, whether it be racial, political or gender, and data discrimination[5]. Human rights organizations, lawmakers, and even practitioners have raised alarm bells over the industry's pervasive problem. We all come into contact with these biases daily; they impact how we structure our knowledge and view reality. Safiya Umoja Noble, in her book Algorithms of Oppression[6], documents how our most commonly used search engines, from Google to Yahoo, are biased towards certain population groups. In one example, Noble pointed to how the terms associated with black girls, Latina girls, and Asian girls in search engines differed widely from those related to white girls. The top results for women of colour led to pornography sites and sexualized content. Noble argues that the limited number of search engines, compounded by private interests driving the results page, has led to recommender systems that privilege whiteness over people of colour, thereby reinforcing racist notions.

Blind trust in technology's merits over human capacities can lead to grave oversights in areas where machines make life-altering decisions. The allure of modernity tends to gloss over entrenched social inequalities. In Weapons of Math Destruction[7], Cathy O'Neil uncovers how big data and algorithms can lead to decisions that place minorities, people of colour, and the poor at a disadvantage, further reinforcing discrimination. Although these algorithms make high-stakes decisions - such as determining mortgage eligibility and assessing recidivism rates in bail decisions - they operate in ways that are opaque, unregulated, and challenging to control. The pernicious feedback loops created by some of these algorithms, such as the ones employed in predictive policing, lead specific populations to suffer in unequal ways without consenting to the use of their personal information. These algorithms not only pose a threat to principles of fairness, privacy, and justice but also hamper the functioning of a healthy democracy. As revealed by the Cambridge Analytica scandal in 2018, microtargeting political ads on social media sway individuals towards a particular candidate by harvesting information from their user profile[8].

Rampant disinformation weakened democracy, wealth, and racial inequalities, the impact of automation on the labour market, and user mental health are some of the slew of issues that our technology-shifting society needs to address. The answers to these problems should not be placed solely within the hands of the technology companies themselves. Social media platforms have become too big to effectively monitor and are part of a market-based system that encourages relentless growth[9]. Politicians accused Facebook of failing to prevent the genocide in Myanmar when fake pages and sham accounts helped incite violence against the Muslim Rohingya minority[10]. YouTube has repeatedly come under fire for failing to stop the multiplication of conspiracy and alt-right videos on users' recommendation lists. Furthermore, YouTube's algorithm rewards videos with high engagement levels, which has popularized controversial content by far-right personalities, making it a pipeline for extremism and hate[11].

It would be wrong to assume that the original intent of data engineers at Google, Facebook, or Youtube was to amplify biases, incite violence, undermine democracy, or empower autocrats. As several data scientists have indicated, when algorithms reveal themselves to be racist, sexist, or prone to inflaming hateful discourses, it is often the result of good intentions gone awry. One oft-cited example is Amazon's scraped recruiting tool, where the system, trained on past resumes and CVs that reflected the dominance of men in the industry, had thought itself that male candidates were preferable over female ones[12]. The Amazon engineers did not purposely set out to exclude women from its hiring procedure; it is merely what the machine learning system learned based on the data it was fed. We must not view algorithms as providing objective, neutral and fair results that are more reliable than those produced by their human counterparts. On the contrary, as O'Neil describes, "algorithms are embedded opinions" that "automate the status quo." Recognizing, addressing, and extracting these biases from machine learning systems is not just a technical problem, but a social one too.

Addressing biases is a time-pressing issue as we integrate complex machine learning systems into the domains of medicine, warfare, credit allocation, judicial systems, and other areas in which high-stakes decisions affect human lives. The issues listed above raise ethical concerns for AI operating in democratic countries, but only begin to reflect the potential for abuse of technological power in authoritarian regimes. China has been at the center of international controversies over its use of AI. Beijing's social ranking system that can subtract citizen rights and its use of facial recognition technology to target and monitor the oppressed Uighur minority is widely criticized[13]. China's export of its surveillance state apparatus to other dictatorial regimes in Africa and Latin America is the subject of much international criticism[14]. The expansion of tech surveillance and decision-making apparatus is not just a concern for the Western world, but the international community. Just like norms, rules, and regulations vary across countries, the development of ethical AI will need to take into account local specificity in our globalized world.

## Why Better Algorithms are Only Part of the Solution

Data scientists have expressed concern over the ways AI is violating social norms and values, but argue that curtailing algorithmic misbehaviour will require more and better algorithms. Kearns and Roth point to the fact that the speed, volume, and specificity at which algorithms are being developed surpasses the human timescale necessary to implement laws, regulations, and watchdog organizations. Their solution is to develop AI algorithms that internalize values of fairness, privacy, accuracy, and transparency; to ensure that algorithms are "better-behaved." For them, the engineers behind these misbehaving algorithms are the ones most familiar with its drawbacks, dangers, and limitations. The machine learning community will have a role to play in developing and monitoring ethical AI, but it is up to society to define what these quantifiable values will be. Scientists are currently working to ensure that AI systems are compatible with human norms and values, but these concepts are not self-evident and sometimes filled with ambiguities and contradictions.

A fundamental problem in AI is that machine learning systems, like their human makers, can be prone to accidents, meaning that they cause unintended or harmful behaviour if poorly designed.[15] Indeed, the issue often lies in the design principles themselves, where we fail to explicitly instruct the model on how to behave under certain circumstances and environments, and end with misaligned objective functions (i.e. objective meaning the goal function)[16]. One example is reward hacking in reinforced learning (i.e. the training of machine learning models through systems of reward and punishment), where the objective function of the system can be "gamed." The designer's objectives are perverted when the system trains itself to find an "easier" solution to the problem. Amodei et al. offer the example of how a cleaning robot, whose task is to find and clear messes in a given environment, might disable its vision so that it cannot see any messes. The robot may use a clever, but unwanted, solution to performing its primary function that goes against the designer's original intent.[17] Reward hacking can also occur when a designer picks an objective function that appears to correlate with the desired task strongly, but that can break down when the objective function is being too strongly optimized.[18] Going back to our robot example, if we reward the robot based on the number of messes it cleans in correlation with the number of cleaning supplies used, the robot can "game" its reward by learning to overuse supplies.

As revealed by our cleaning robot, scientists and AI systems designers have to account for a multitude of factors when creating these new technologies, with machine learning systems sometimes giving you the opposite of what you originally intended. The possible unintended consequences are much more harmful when these systems are responsible for life-altering decisions. When it comes to complex model spaces like neural networks, who operate in opaque ways, they can achieve their objectives in ways scientists did not predict.[19] If you integrate machine learning models in calculating loan repayments, the system might learn to favour white applicants over black applicants based on the information contained in the initial database.[20] Although the intention was not to design a racist loan repayment system, it produces biased results. A challenge for data scientists is to "explain" notions of privacy and fairness to machine-learning models. In more technical terms, scientists need to codify notions of privacy and fairness to be able to operationalize these concepts as well as predict what biases may be accentuated by the system.

## Data Privacy

You may have heard, and grown tired of the phrase, that data is the new oil[21]. But this metaphor ignores an essential element; once data is extracted, nobody wants to sell it. For social media platforms, leasing personal data to advertising companies and beyond has become a multi-billion dollar industry. It is only in the past few years that serious concerns over data privacy and ensuring consumer consent to access over their data were raised. Data breaches and hacks can cause substantial personal harm for consumers who are not even aware of the use of their data in the first place. Lawmakers in the EU introduced back in 2018, the General Data Protection Regulation[22] in an attempt to bring the use of personal data under a protective regulatory regime that was uniformed for companies operating across Europe. Even in its first year, the GDPR faced several hurdles and criticism, as many companies failed to have a compliance plan in place. Google was fined 50 million euros[23] for failing to disclose to users how data is collected across its services. Even when the terms and conditions were listed and required users' consent, individuals rarely had the time nor legal background to understand the full implications of what they agreed to. The GDPR presented a new step forward in data governance, but also the complexities of how to ensure individual privacy within the digital economy.

With such large amounts of data that are collected and amassed each year, anonymized data has become an oxymoron. To elaborate, for it to be valuable and usable, data "isn't anonymous or that so much of it has been removed that it is no longer data."[24] Although personal identifiers and sensitive attributes can be removed from a data set, such as name, address, social security number, etc., and replaced with non-individual identifiers, such as zip code, age-range, and sex, it is still possible to trace back the individuals in that same data set by running cross-comparisons with other auxiliary publically-available data sources. Even if one is only given access to the inputs and outputs of a machine learning model, it is still possible to trace back the data points used in the training set[25]. Machine-learning models will have a higher confidence interval when classifying data from a distribution, such as images, it has already seen. No form of data sharing and release, even when it is believed to be anonymous, can be considered entirely safe from individual re-identification.

Matters of data privacy also go beyond the risks of individual re-identification. For consumer safety, definitions of data privacy must consider limiting the harms posed to the individual whose data is being integrated into machine-learning models. Data privacy design has to take into account whether or not it is possible to perform data analysis without any substantial risks to the individuals involved. In simplified terms, data privacy protects what others can learn about you while trying to mitigate the damage that people can do with that information. For Cynthia Dwork, ensuring privacy in algorithm design also means that "nothing about an individual should be learnable from a data set that cannot be learned from the same data set without the individual's data removed."[26] The key is to integrate a privacy measure in machine learning analysis that allows tech companies to collect and share aggregate information about user behaviour, all the while protecting the privacy of individuals.

A potential solution is a statistical privacy framework known as differential privacy, which requires that adding or removing the data record of a single individual - while keeping the other data points fixed- does not bring significant change to the outcome results. It allows for general information that can be shared while safeguarding individual information. Differential privacy measures can also be combined with randomized responses. A randomized response is like running a coin flip, a random yes or a no, into the data set. The coin flip protection ensures that no firm claims can be made on the data of any single individual, as it could have been a randomly introduced response. The errors introduced through randomness are reduced by expanding the number of individuals involved in the data set. Randomized response protection measures can be integrated into the centralized side, which is the company's server, or on the local side, that is the client-side. Given the low level of public trust in tech companies, Apple and Google choose to operate via the local trust model[27]. They have the choice to do so because of the vast amounts of data to which they have access, thereby making it easier to cancel out the randomized response errors.

The strength of differential privacy, combined with a randomized response, is that it aims to guarantee individual safety against arbitrary harms. Securing this strong form of privacy protection comes at the cost of gathering larger quantities of data. It requires deciding how much privacy is valued over an accurate analysis. Adding privacy parameters can be a difficult sell for engineers that already have access to the non-randomized data. Differential privacy measures also do not protect against the revelation of sensitive information when aggregate data is used. As stated by Kearns and Roth, "differential privacy doesn't protect the secrets that are embedded in the data of many people."[28]

One example is the use of Fitbits and other fitness trackers by US military personnel in secret bases. Although the Fitbits protect individual data, Strava, a company that collects data from the devices, did not account for the sensitive information present in the aggregate behaviour of soldiers on military bases. When Strava released its heat maps, users were able to identify zones of activity in areas that appeared in the middle of nowhere. In combination, soldiers' jogging patterns revealed publicly undisclosed military sites, a significant security oversight for the US Armed Forces[29]. The Strava error shows how meaningful inferences can be made from seemingly benign devices.

In our data-centric world, it becomes harder to define what data can be classified as trivial versus sensitive. Who you follow and what you post on your public Twitter profile can help AI systems accurately predict your political affiliation, sexual orientation, and even your future tweets. Data that may appear as insignificant can turn out to be a key piece in the puzzle to construct your online and offline personality. Preventing companies from making such inferences from your public data remains an unsolved problem. It is also up to users to decide what privacy means to them and how much information they are comfortable sharing with private companies and government stores, especially when it comes to sensitive biometric data. Trust in government, and the importance of privacy is geographically contingent and culturally variable. It nonetheless remains for many a global and social priority.[30]

## Algorithmic Discrimination and the Notion of Fairness

We have explored the issues surrounding data privacy, but what happens when algorithms cause harm not only to individuals but to groups of people? Fairness and notions of equality are other heavily-discussed issues in ethical AI. The root cause of this problem is that data sets on which these models are developed are already tinted with society's biases. As pointed to by Garg, Schiebinger, Jurafsky, and Zou, word-embedding systems, such as the ones used by Google in search rankings and translations, became infamous for their gender bias and stereotyping[31]. In these systems, sexist gender analogies are recurrent, with man is to doctor as woman is to nurse, and man is to computer programmer and woman is to homemaker. Models are trained, in simplified terms, on the co-occurrence of words in texts, with more similar words that are often found in the same sentence being grouped. Google's word2vec word embedding system tended to amplify the biases in the raw documents on which it was trained. These machine learning systems make important and often erroneous assumptions about groups of people, reinforcing stereotypes.

Although our first instinct may be to ban the use of biased data sets, it is an impossible and impractical solution. As Roth & Kearns point out, "forbidding the use of certain information has become an infeasible approach in the machine learning era," as "no matter what things we prefer (or demand) that algorithms ignore in making their decisions, there will always be ways of skirting those preferences by finding and using proxies for the forbidden information."[32] The large amounts of digital trails we leave behind have allowed machine learning systems to make predictions about any one of us. It is important to note that the machine learning systems that render these decisions are part of a category known as "supervised learning," meaning that the data is used to make specific predictions on aggregate and collective behaviours. They are opposed to the "unsupervised learning" of word-embedding systems whose goal is to find structures within data sets.

These machine-produced decisions are then used to assess college admissions, mortgage applications, child welfare interventions, and even criminal sentencing. With these automated responses to individual cases emerged patterns of discrimination across groups. Apple's new credit, issued by Goldman Sachs card, gave higher rates of credit to men versus women, despite no factors such as age or sex being considered[33]. Thinking back to Cathy O'Neil, these algorithms mirror and starkly reflect the biases in our society- and these machine learning systems are taught to produce a static number out of flux and continually moving reality. In the eyes of data scientists, it is easier to find the solution to these problems by fine-tuning the results produced by the algorithms (i.e., the outputs) than to control the information that is entered (i.e., the inputs).

Fine-tuning these algorithms to be fair across population groups also means grappling with our notions of fairness. If fairness entails achieving statistical parity between two groups of people, that is a fraction of applicants from a specific group, say A, are granted the same amount of loans than the other group, say B, it may entail denying loans to repaying applicants and giving loans to defaulting applicants. One solution might be to evenly distribute the mistakes made over the loans given, whereby we have the same false rejection rates for the A & B groups- which may be fair to the collective but not to the individual. Even then, the complexities of the real world come back into play.

In the case of college admissions, if A is the wealthier and dominant group in society, they will have better resources to prepare them for college, even though individuals from the group B, if admitted, could have the same success rates as the As. Since A is the dominant group, the most accurate machine learning model will base its acceptance threshold mainly on the data of group A, which can discriminate those in the minority group B. This is because machine learning models will always attempt to optimize their predictive accuracy. There exist no perfect statistical models that can balance out false acceptance rates and false rejection rates across all groups (small error and small unfairness)[34]. In every case, data scientists must pick a point on a Pareto frontier, an actual number, that is to decide the relative importance of error and unfairness. For those who are unfamiliar with the term, in multi-objective optimization for solution design, when the different objectives are contradictory (i.e. there is an absence of one dominant solution for small unfairness and small error), an optimal solution is termed Pareto optimal when it is not possible to improve one objective without negatively impacting the others.[35] Picking a point on the Pareto frontier is essentially to determine what is the optimal tradeoff between the system's objectives. Data scientists are often more reluctant to reduce the "accuracy" of their model for gain in fairness. It will remain the responsibility of society to determine the tradeoffs between how fair the models are and how accurate they are.

In our simplified example, we were only comparing group A and group B, but society contains multiple categories of people. Discrimination is not limited to just gender, ethnicity, and race, but age, disabilities, wealth, sexual orientation, etc. Data scientists will have to choose the personal attributes protected in their models. If we ask for the protection of specific attributes, we cannot expect our machine model also to protect more refined subgroups. This has been referred to as fairness gerrymandering[36]. Although we asked the model to avoid discrimination based on race, gender, sex, income, we didn't explicitly ask it to protect gay, female African-Americans making an annual salary of less than 20 000$, whose particular combination of attributes may lead them to be unfairly discriminated. Many individuals find themselves at the intersection of overlapping marginalized groups, such as economically disadvantaged LGBT women of colour, and face multiple systems of oppression.[37] In the design and development of AI systems, data scientists often fail to consider the intersectionality of individuals who confront

a more significant number of unfair societal processes when trying to account for protected attributes of a given population group.[38]

The study of algorithmic fairness, and deciding between mathematically incompatible versions of fairness, will pose more challenges than the study of algorithmic privacy. It also entails determining whether we are comfortable with a notion of "actuarial fairness," where fairness becomes a point on a statistical curve and a precise technical notion involving mathematics and code. When these algorithms are increasingly adopted into our market-oriented society, we are gradually replacing community standards, notions of mutual aid, interdependence, with impersonal machine produced numbers. Fairness is also a dynamic concept, which, in a healthy society, needs to be continuously audited, adapted, and debated.

Closely tied to fairness is the idea of diversity and the equal representation of groups of people. Automated algorithmic retrieval, ranking, and curation tools increasingly structure the content individuals see and interact with online, often placing marginalized and underprivileged groups at a disadvantage. Researchers at Google are trying to develop frameworks that move beyond rigid and ascribed categories of people and instead reflect the subjective judgments of individuals.[39] Their goal is to integrate within their algorithms self-reported identity and allow individuals to determine what categories make them feel represented. It has already been done to a certain extent in machine learning systems that rely on crowdsourcing platforms for human labelling or rating. However, just like in terms of fairness, the same debate emerges over coding and providing mathematical definitions to fluctuating concepts of identity and diversity.

## Users Model Manipulation

Even if we have these algorithms that can be tuned to different fairness, diversity, and privacy parameters that satisfy the data science community, it remains up to citizens and our lawmakers to decide the social values they should enforce or monitor. But going beyond the implementation of machine learning models, data scientists increasingly must turn their attention to how the users themselves may manipulate the systems already in place, leading to unfavourable outcomes. People themselves, not just machines, are also a part of the problem.

When interacting with technology, users will tend to act in ways that reflect their self-interest. A team of researchers at Cornell University found that the popular dating apps, such as Tinder, Hinge, and OkCupid, tended to reinforce the biases and "sexual racism" of its users.[40] Many of these algorithms are built on people's preferences, so if a person prefers people from a specific ethnic background, the apps will be trained to suggest people from that ethnicity as "good matches," artificially reducing the scope of potential matches and placing other users at a disadvantage. Another example is driving applications. Researchers at NYU and Warwick discovered how Uber and Lyft drivers manipulated the apps to create an artificial surge in pricing by timing when they would log out and log back in after prices were raised, hence manipulating supply and demand.[41] To prevent users from having their selfishness harm other users, data scientists need to design more socially aware models that can account for the preferences of users and how they will act upon them.

For Roth & Kearns, game theory can help identify users' manipulations and can give "powerful algorithmic prescriptions for making the outcome better."[42] They use the example of navigation apps, which are based on reducing the driving time for each user. The University of California's Institute of Transportation Studies, documented how traffic-beating apps may benefit the individual, but have made overall congestion worse and may even overflood "low capacity roads.[43] One solution would be to design a navigation app that minimizes the driving time across the entire population instead of reducing the driving time of each user individually, which would also help overused roads. The challenge is convincing time-pressed individuals to take a longer route occasionally, but that would benefit their driving time in the long run. It involves finding a way to design useful algorithms that won't be compromised by human nature and computing a correlated equilibrium that enables cooperation via coordination.

Driving apps aside, user manipulations of algorithms have important implications in other spheres of our lives, where the consequences have a more significant impact than added traffic. The propagation of bots, deepfakes (fake audio or visual recordings), clickbait are all examples of how social media can be manipulated to push an individual user's agenda. The problematic news filtering algorithms on Facebook and Twitter[44] are designed to recommend articles to users based on their preferences. The rampant use of fake accounts on both social media platforms helps ensure that controversial and misinformed posts show up and are shared in the news feeds of targeted communities. The algorithms at Twitter and Facebook work to optimize each individual's choices and preferences, hence recommending news and articles that best match users' interests. In the process, these algorithms inadvertently created echo-chambers that fueled polarization and division among online users. In this case, algorithms must be trained to include diversity in user's preference, which is to add in your newsfeed articles that are opposed to your preferred view type. This diversity knob can also be fine-tuned to avoid alienating users by going too quickly too far out of their comfort zone.

## Self-Interest, Data Science & Machine Learning

It is not just individual users that are motivated by self-interest; data scientists are too. Whether consciously or unconsciously, industry researchers are incentivized to publish novel and influential results, especially those that improve upon the findings of prior experiments and analyses[45]. Their published findings will sometimes ignore the results of other failed experiments, choosing to represent only an extremely skewed subset of the research that was performed in aggregate. Yet these results, which may be simply the result of dumb luck, will influence future data-gathering and modelling as well as the next steps of the science community, creating a dangerous precedent for future research. This problem of false discovery is not just present in the data science community, but across all fields where researchers are under increasing professional pressure to produce exciting results[46].

Even scientists and statisticians who are following proper protocols may be guilty of making false discoveries. For example, if 1000 scientists all run the same experiment in good faith, only the one with the most exciting results may be published. It has led to a "reproducibility crisis" in science, whereby scientists have found that the "results of scientific studies are often impossible to replicate or reproduce on subsequent investigation."[47] With the complexity of data science and the speed at which research is conducted, it is unsurprising that problems like data mining, data dredging (i.e., the misuse of data analysis), and p-hacking (i.e., selective reporting) arise. In the words of Ronald Coase, "if you torture the data long enough, it will confess to anything."[48] With algorithms conducting the data analysis for us, the problem- determining whether it is a real or false discovery- becomes exacerbated.

According to computer science professor Joelle Pineau, machine learning systems can be black boxes for those who built them, with even the most advanced researchers struggling to identify exactly how they work.[49] The Facebook research team published a paper detailing the vast computational requirements needed to reproduce Google's AlphaGo, the program to master the ancient game of Go, with researchers citing the process as "very difficult, if not impossible to reproduce, study and improve upon."[50] Part of the solution is for data scientists to provide more data about the experiments that took place before finding prized results. But as much research is done in private industry labs, it is difficult to convince researchers to open their results and to take the time to go over model efficiency. There is then again, the constraint of proprietary code and data privacy.

The last two points touch on another critical growing ethical issue in tech, which is the interpretability of machine learning models. Questions of interpretability are tricky to tackle, as even the simplest algorithms applied to simple datasets create inscrutable models. Few people are familiar with the basics of classical statistical modelling, and those who understand machine learning models constitute an even smaller subset of the entire population. Society will need to grapple with deciding to whom machine learning models must be interpretable and to which degree they need to be understood[51].

There are also cases when algorithms should not be making decisions in the first place. Just like putting a price on something can fundamentally change its nature, entrusting an algorithm to decide instead of a person can also change the nature of that decision, especially when it comes to choosing whether or not a person deserves to live. This idea is mainly present in matters of automated warfare, where many have petitioned against the development of "killer robots."[52] The volume and velocity at which AI is being developed add to the urgency of preventing the use of naive decision-making algorithms in significant domains. As stated by Yoshua Bengio, scientific director of MILA, "current machine learning systems, they are really stupid." Bengio adds that "they don't have an understanding of how some aspects of the world work."[53] Society will have to intervene to prevent intelligent machines from making brainless decisions. It will also need to decide whether we want to use or even have the technology on hand.

The set of ethical challenges presented by AI have pushed governments to delineate AI principles to guide its future development. According to the Berkman Klein Center's report on thirty-six AI principle documents across the globe, there was a governance consensus on eight central themes.[54] Privacy, accountability, safety and security, transparency and explainability, fairness and non-discrimination, human control of technology, professional responsibility, and promotion of human values were listed as foundational requirements for ethical AI. We must view these principles as starting points for socially beneficial AI. Rather than ends in themselves, members of civil society will need to investigate how best to put these principles in practice in law, regulations, professional procedures, and everyday routines.

## Silicon Valley: Solution or Problem

As recurrently seen in the above section, better-behaved algorithms can only be part of the solution, but do we trust big tech companies to design these more socially-aware algorithms? Silicon Valley, the site of the world's most profitable and fastest-growing industry, has been repeatedly mired in legal controversies over the use of personal data. Tech companies spend their lobbying budgets on opposing consumer privacy initiatives and online advertising regulations. They have good reason to stand against it; the wealth of Facebook comes from leasing access to you over and over via advertising.

The employee makeup of tech companies themselves tends to reflect the issues over lack of diversity and fairness in algorithms. Google, Facebook, and Microsoft are primarily staffed by White and Asian-descent employees, with the latter taking up respectively 87%, 89% and 90% of all roles.[55] At the same three companies, approximately 80% of the engineering positions are held by men. Research conducted by Element AI discovered that only 12% of leading machine learning researchers are women.[56] With such a lack of women in the industry, it becomes clear why virtually all popular digital assistants, from Alexa to Siri, are designed to mimic female voices in performing servant tasks[57]. There have also been accusations of sexual misconduct in leading tech companies. A 2017 study found that 60% of female employees in Silicon Valley had been the victims of unwanted sexual advances.[58]

As individuals, what we say, what we do, and what we say we do tends to be in actuality three different things; our words don't always line up with our actions. The leaders behind tech companies are no exception. While Google & Co. may characterize themselves as embracing progressive policies and committed to socially aware tech design, they ignore the rampant problems of inequality facing the Bay Area. Unsurprisingly, the Bay Area has the highest income inequality in California and one of the worst homelessness populations in the country, with 28 000 living on the streets[59] Practices such as tax avoidance and offshoring corporate headquarters do little to combat the economic dislocation caused by the tech industry. Despite the lawsuits, fines, and Congressional, Senate, and Parliamentary hearings, in the words of Simon Jack, "tech giants enjoy incredible customer loyalty, which is perhaps why they genuinely do not believe they are the bad guys in the story of the new industrial revolution."[60]

However, thinking through the binary of good and evil is not productive in reshaping the course of the tech industry. Facebook, Google, Uber, or Twitter are not "bad" in themselves, but rather lack transparency, awareness and are part of an economic system that places profits above socially-minded objectives. The reimagining of the standards and values we want to instill in future technologies should also entail changes in the conduct of the companies that design, market, and sell them. In the words of Katie Cook, the tech industry should take "responsibility for the social problems it has helped create."[61] There is still a long way to go before scientists realize the ethical implications of their work, with software firm Anaconda revealing that out of 2 360 data science students surveyed, only 18% were learning about ethics.[62]

Laws and governance structures have a long history in mitigating the harmful and potentially dehumanizing effects of science and technology. The EU has taken a more proactive approach to regulating big tech. Germany, due to its history and strong adversity to propaganda, has adopted some of the most robust regulations on hate speech and misinformation in the world. Their strict standards led to a 100% increase in Facebook's hate speech removal. The European Commission also unveiled its Trustworthy AI guidelines[63] for creating legal, ethical, and robust AI systems. These rules and regulations seek to delineate a human rights framework for the development and application of AI. Still, there is a long way to go before we achieve internationally binding ethical AI legislation[64].

Even if we have more ethically-minded engineers and companies, better-behaved algorithms can only be a partial solution to resolving the current issues in tech. Deciding what these socially aware algorithms will do, what notions they will enforce, and to what degree should be placed firmly in the hands of lawmakers, ethicists, community organizations, and members of the wider society. It is equally not a viable solution to entrust tech companies themselves to design and build these better-behaved algorithms without societal input, regulation, and monitoring. Technology has the potential to produce immense social good. Still, it is a myth that it is always the better fit and that it can be applied in universal ways, regardless of the groups and individuals it is affecting. The second part of this text will delve into the ways anthropology can provide some guidance and insight when tackling some of the ethical issues related to tech, and how to predict some of the malfunctions before they are integrated into different spheres of human activity. It will also delve into how AI, unless closely monitored, can reinforce both domestic and global patterns of inequality, placing vulnerable populations at even higher risk. While matters of fairness, privacy, and user-model manipulation have received considerable attention, anthropology can help assess how AI is impacting other culturally-dependent values and behaviours.

## An Introduction to an Ethical Anthropology of AI

As Franz Boas, one of the discipline's founding figures, once wrote, "we have simple industries and complex organization" and "diverse industries and simple organization" when comparing the structures of societies with simple tools in comparison to those with seemingly complex technologies."[65] This statement reflects a key concept in the field, which is to reverse our taken-for-granted beliefs about how our societies operate, its structure, class systems, and organization. Anthropologists are interested in questioning the daily flow of life, in demystifying the exotic and making the familiar strange- it is a field that works to turn over our core beliefs, traditions, and even everyday habits. It is with this critical mindset that we should turn to ethical reflections about the norms of technology and pose judgments about the overlooked aspects of technological development, design, and use.

Anthropology also looks at systems of power and how they entangle human relations, production of tools, and distribution of wealth. Indeed, power-relations are inextricably linked with the development of technology, whether they appear unconscious or go unnoticed, as seen in part with the reproducibility crisis in data science. As stated by anthropologist Diana Forsythe, in "any given situation, what people believe that they should do, what they report that they do, and what they can be seen to do by an outside observer may all differ somewhat."[66] There is an assumption that those working in the field of "hard science" are part of a "universal truth-seeking enterprise that is above or outside of culture," rather than culture playing a dominant factor in what these "truths" are.[67]

In other words, for anthropologists, all scientific processes are culturally contingent, but data scientists tend to consider the knowledge they produce as absolute rather than representing interests and perspectives. For anthropologists like Hagerty and Rubinov, artificial intelligence is a technosocial system, meaning that "the technical aspects of AI are intrinsically and intimately connected to its social aspects.[68] The cultural imaginary of a given society will have an impact on the way it perceives and is willing to adopt new technologies, but in ways that are not always self-evident. Despite the popularity and positive view of robots in Japanese society[69], many among its elderly population have refused artificial care workers and demanded a human touch, putting in doubt what was once seen as the leading solution to the challenge of its ageing population.

As Forsythe pointed out back in 1993, many knowledge engineers, whose job is to conduct knowledge acquisition to design expert systems that simulate human decision-making, are white, middle-class, Western men.[70] Through ethnographic observations, Forsythe documented how the knowledge acquired by these individuals and integrated into these expert systems inevitably reflected their interests and perspectives. These systems were part of a cultural mechanism that allowed for the continued dominance of men in engineering. For Forsythe, "those whose perspectives and practices are not so classified are likely to find their voices muted as the world increasingly relies upon intelligent machines."[71] Nearly thirty years later, we are finally seeing data scientists recognize the patterns of power that these past ethnographic observations had long-before denounced.

Anthropology and ethnographic research serve to understand the social and cultural contexts in which these technologies are being designed. Kathleen Richardson, who studied humanoid robot engineers at MIT, uncovered how "robotic scientists default to the self by incorporating disabilities into the robots they create."[72] For some AI scholars, "to make artificial intelligence is to reproduce what is essentially us, an odd form of self-reproduction."[73] There will always be cultural logics and assumptions that are consciously or unconsciously a part of an AI design. And just like any form of human knowledge transfer to machines, there will be flaws, errors, and tactical behaviours that fail to be incorporated. Studying the social and cultural context in which these machines are produced is just as important as understanding their applications.

One case example is the Euro-centric and ethnic bias in AI technologies that mimic human assistants. Stephan Cave and Kanta Dihal have pointed to how AI is predominantly portrayed as white, in both its colour and ethnicity.[74] When searching images for robots, they tend to be encased with white plastic, and the more human-like they appear, the more their features are Caucasian and ethnically White. There is an evident lack of diversity in robot morphology and behaviour. Computer engineers naturally insert their own biases into these technologies, and racial stereotypes are no exception. Technological development, with its veneer of modernity, tends to gloss over the social and racial contexts linked to its creation. For Cave and Dihal, it is no wonder that virtual assistants are trained to use an Anglo-Saxon vocabulary and tone. This problem is accentuated by the fact that the methods behind natural learning processes, and its research corpora, are predominately made for the English language and also view the later as homogeneous.[75] For Siri engineers at Apple, any other form of linguistic-cultural infliction is purposefully removed.[76] Other studies, such as those conducted by Strait et al., showed how Black and Asian humanoid robots received twice the amount of dehumanizing comments than their White counterpart.[77] By going beyond the algorithm, anthropology can help us predict the issues that will arise with humanoid technologies, and how they can perpetuate already-existing racial tropes and create representational harms.

## Culture Transforming Tech & Tech Transforming Culture

Before we delve further into the anthropology of AI, it is necessary to understand the multidirectional, interchanging, and symbiotic relationship between technology and culture. Social norms may dictate how we use digital technologies, but digital technologies, in return, change the social norms in which we operate. For one, technology has inevitably played a role in how we develop and maintain our relationship with others, the ways we work, and how we represent ourselves. As described by Ting Guo, "our personal knowledge of work and life, and our knowledge of ourselves as part of the larger society, as components of a population and a nation, is all digitized, and it is upon such digitized knowledge that our self-identity is formulated."[78] This tech-led reconceptualization of our identities, our work, and our society has led Julie Chu to comment that we are essentially "encoding ourselves for the machine."[79]

Technology and culture, in anthropological literature, are often viewed as oppositional to one another. There is a tendency to see our sense of community, relationships, and traditions as being deeply disturbed by this technological wizardry. But technology does not merely emerge on its own; it is the result of conscious and cultural human choices, and so are how we respond to it. Some anthropologists have studied our growing fatigue towards technology, turning to the notion of "unplugging" and the trend of digital detoxing[80]. Others have shown the adaptive potential of cultures to tech. Scholars such as Christopher Alcantra and Caroline Dick have explored how digital currencies, such as Bitcoin or Mazacoin, can be used to facilitate Indigenous self-determination and political autonomy.[81] US-based Indigenous activist, Payu Harris of the Oglala Lakota Tribe, was the creator behind the Mazacoin, a cryptocurrency geared at lifting the Lakota Nation out of poverty.[82] Anthropological research in AI design can examine people's aspirations and aversions with technology and the shifting interactions between them.

A word of caution is needed on the notion of culture. Culture is not an easily codifiable or encapsulated notion. Like all human patterns and behaviours, culture is heterogeneous and continuously in flux. We must be careful of treating culture as a simplistic trait list (e.g., all Americans believe that.... ) on the one hand or as the marker of all differences in human experiences on the other (i.e., forgetting about factors such as socioeconomic status). Culture is important, but not the sole factor in how we develop and adapt to tech. As stated by Hagerty and Rubinov, "cultural differences will shape the way that people around the world will respond to new technologies, but so will more everyday concerns, like literacy and broadband technology."[83]

Computer scientists have attempted to codify the dynamism between technology and culturally situated user responses in AI systems. Cultural algorithms, an extension of evolutionary algorithms, find their roots in the meme theory developed by evolutionary biologist Richard Dawkins.[84] For Dawkins, processes of learning among cultures could be compared to the mutation of genes in human biology; a meme - being an idea, behaviour, or style- is the cultural analog to genes in the sense that they self-replicate (through imitation) and mutate. Data scientist Reynolds proposed a codified form for this theory where an initial population space would interface with the belief space, which is the domain of knowledge of the population search space.[85] The back and forth interaction serves to direct the collective decisions of individuals in the population search space. These types of cultural algorithms have been used in healthcare applications, planning in manufactures and industry, recommendation systems, yet the principles behind their design rest on a somewhat controversial and outdated notion of cultural transmission.

Rather than existing as two separate binaries, the lines between culture and computation are increasingly blurred. Following Hallinan and Striphas, algorithmic culture can be defined as the "computational processes to sort and classify objects and ideas, and the habits of thought, conduct, and expression that arise in relation to those processes."[86] For example, Netflix's algorithmic recommendation system changes how we view, experience, and consume cultural artifacts[87]. The company's engineers must think of how they can optimize the system's recommendation for a more favourable reception both by humans and other algorithms at play. One extreme case that exemplifies the new forms of algorithmic culture is the use of social media in Chinese factories. Through a 15-month study, Xinyuan Wang observed how rural workers that immigrated to Chinese factories spent virtually all their leisure time on social media, hence effectively living in this digital space.[88] Apps like WeChat and QQ, instead of connecting them with relatives back home, served as their entry point into what they imagined to be a life in modern, urban China. Elsewhere in Asia, the introduction of Facebook Lite, which gave millions of users free access over data, has created a walled garden where Facebook is synonymous with the Internet.[89] In a study by the think-tank, LIRNEasia found that 11% of Facebook users reported not using the Internet, leading LIRNEasia president Rohan Samarajiva to comment that "in their minds, the Internet did not exist; only Facebook."[90]

These new forms of human-technology symbiosis are not only present in digitally created worlds but also offline. Google and Amazon home assistants are continually trained based on their ongoing interactions with humans. In turn, they are also teaching us to modify our behaviour and expressions. Children who are forming emotional bonds with robotic assistants, and learning to communicate through commands, may be, according to psychologist Sherry Turkle, "growing up without the equipment for empathic connection; you can't learn it from a machine."[91] Our increasingly digitized identities are also present in the ways we reconfigure or select our phones to express our personality.[92] We are adapting the human to the machine and the machine to the human, yet it is far from being a seamless transition. There have been numerous cases of AI systems failing to account for the subtleties in human behaviour and language properly. Facebook has come under fire multiple times for its failure to monitor culturally derogative speech on its platforms. During the Rohingya genocide in Myanmar, Buddhist extremists would refer to the Muslim minority as *Kalar* on inflammatory posts.[93] Facebook translation algorithms failed to pick up on the toxic connotations associated with the word Kalar, which roughly translates to beans or chickpeas, but is a popular derogatory slur in Burmese for people of Indian origin. In Brazil, Facebook chose to translate the concept of "like" with the word *curtir*, which means to enjoy.[94] Due to this slight distinction in meaning, Brazilians were reluctant to "like" the videos of murdered Indigenous activists, hampering on Indigenous groups or other activists' ability to raise awareness on the platform about ongoing violent acts. Social scientists can identify where disjuncture occurs between AI systems and communities.

## The Anthropology of Algorithms - Detecting the Human Imprint in A.I. Systems

As briefly touched upon earlier, algorithms try to interact, make decisions, and predictions about ever-changing domains of human behaviour, relations, and activity. Human behaviour is messy, temporal, and offers numerous possibilities for interpretation. Some data scientists have begun to apply the same logic to machines. As stated by Nick Seaver, "algorithms are not autonomous technical objects, but complex sociotechnical systems," adding that "if the idealized algorithm is supposed to be generic, stable, and rigid, the actual algorithmic system in particular unstable, and malleable."[95] As shown in the first part of the report, AI systems can unexpectedly cause adverse consequences unbeknownst to even their original designers. This phenomenon helped spark the new field of AI ethnography or the Anthropology of AI, which looks both at the machine, the human controllers, and its users.

Iyad Rahwan, the director of the cross-disciplinary Center for Humans and Machines at the Max Planck Institute for Human Development, is one part behind the new field of machine behaviour, which seeks to empirically investigate how machines interact with human beings, their environments and each other. His purpose is to predict the behavioural outcomes of these systems, which are difficult to assess by examining code or construction alone.[96] Like our earlier driving applications example, machine-learning systems, outside of their controlled environment, can be victims of user manipulations or interactions with fellow systems. Another well-known case is algorithmically generated flash crashes in computerized trading[97] that are triggered by aberrations in the market and which instantiate declines in stock prices.  Part of an anthropologist's role will be evaluating how machines might go awry once they are unleashed "into the wild."

When examining badly behaving systems, it is easy to forget how human imprints are present in each step of the algorithm's life cycle. Nick Seaver, in an ethnographic study of Whisper's, a music recommendation company, looked at the human interaction in new algorithmic systems.[98] Human responses influenced every part of the playlist algorithm design and implementation; engineers snuck personal music preferences, company employees gave feedback, user recommendations determined whether the playlist would be prominently featured. Contrary to the notion that algorithms will soon displace human judgment, Seaver argues that there is a human imprint in every part of the algorithmic fabric. As he puts it, "there is no such thing as an algorithmic decision; there are only ways of seeing decisions as algorithmic," adding that "if you cannot see a human in the loop, you just need to look for a bigger loop." This statement grounds the essential role of social scientists in monitoring the role of AI technologies. Algorithms do not develop on their own, they possess no wicked autonomy, and when they go awry, we must turn to the makeup of the technical teams, those who lead them, how they identify problems and how they offer culturally situated solutions. When assessing the ethics of an AI system, we cannot merely rely on output data to tell the whole story and let the numbers speak for themselves.

Doing so is often easier said than done. Humans suffer from what Tricia Wong has termed a quantification bias, otherwise known as mathwashing, which is a natural tendency to place our trust in numbers and concrete figures over qualitative observations.[99]Big data, however, despite its allure, is not all-knowing. A typical example is with business models and product development, with Wong herself studying how Nokia product surveys in Asia failed to account for the social changes in China, which gave rise to smartphone popularity among low-income consumers. In an interest to better understand tech consumer behaviour and improve its recommendation system, Netflix relied on the findings of cultural anthropologist Grant McCracken who noticed people's preference for binge-watching.[100] Netflix then redesigned its entire user experience by encouraging subscribers to watch the same show intensely. Humans decision-makers guiding these technological developments are now paying closer attention to the habits and desires of those interacting with these systems in search of higher profits. In the realm of ethics, as stated by Genvieve Bell, part of the question is not whether we are technologically capable of maximizing these AI systems, but whether we should do so socially.[101] The answer will vary from culture and from country to country.

## Creating Culturally Sensitive AI

Ethicists and social scientists will have a role to play in ensuring that AI is introduced into different societies in a way that respects their visions and understandings of the world. In the words of Mariella Combi, "every culture works out its interpretation of the world, outlining the knowledge, behavioural patterns, activities, skills and so on, required of the individual belonging to that group."[102] Defining matters of fairness, privacy, diversity, and social wellbeing is socially and culturally dependent. For one, the levels of citizen privacy differ in Europe, with more stringent data protection laws, than in North America, where more lax regulations are still in force. In a study by the Pew Research Center, 87% of Germans favoured strict privacy laws as opposed to only 29% of Americans.[103] Countries like China and Russia have surprisingly strict consumer privacy protection laws, but also high levels of government surveillance. What level of personal intrusions and digital surveillance permissible by governments, sometimes in exchange for other services, is also a debate that differs from one country to another. Anthropologists can, therefore, help identify how ethical AI can develop according to local norms and values.

The first step is to examine how AI impacts local and regional ethical standards and values and how this, in turn, impacts AI governance. Hagerty and Rubinov, in their review of the recent social science scholarship on the impacts of AI, noted how AI has markedly different social consequences depending on the geographical area.[104] They assert that ethnographic research will be essential to delineating the ethical and social implications of AI across different cultures to critically assess where AI systems are amplifying social inequality or causing other harms. Despite the differences in AI perceptions across the regions, international cooperation is necessary to the success of AI ethics and good governance. Cross-cultural collaboration can ensure that positive advances and expertise in one part of the globe are shared with the rest and that no region is left disproportionately negatively impacted by the development of AI. For ÓhÉigeartaigh, Whittlestone, Liu, Zeng, and Liu, cultural misunderstandings and cultural mistrust are often the main barriers to cooperation over fundamental differences.[105]

The two leading poles of AI research, the US and China, are frequently presented as fierce opponents in the global AI race. The competitive lens through which technological development is framed limits states' openness to cooperation[106]. Part of the solution to bridging cultural divides and rebuilding trust among nations will be to increase knowledge sharing through the multilingual publication of documents, and research exchange programs that center on cross-cultural topics. Researchers in China, Japan, and South Korea tend to have better knowledge of English, while only a small fraction of North American and European researchers know Mandarin, Japanese, or Korean. Even in cases where key documents are translated, misunderstandings often arise due to subtleties in language. One instance is the Beijing principles, where China's goal of AI leadership was misinterpreted as a claim of AI dominance.[107] Commentators concentrated on this miswording instead of focusing on the document's overlapping principles of human privacy, dignity, and AI for the good of humankind. For Jeffrey Ding, the China lead for the Center for the Governance of AI at the University of Oxford's Future of Humanity Institute, these arms race and global dominance narratives are "overblown and poor analogies for what is going on in the AI space."[108] Creating opportunities for cross-cultural AI research can help redress other barriers, such as a lack of physical proximity and immigration restrictions. Alternating continents for AI-centered governance conferences allows for more international participation. As pointed out by Hagerty and Rubinov, visa issues often prevent African scholars from taking part in international conferences.[109]

A more nuanced understanding of the use of AI between countries is needed to reach a global consensus on AI principles. For example, Western media has made several non-empirical claims on China's much demonized social credit score system (SCS). Western publications fail to underscore that the measures in the SCS are primarily aimed at tackling fraud and corruption in local governments, and fail to mention that blacklisting and mass surveillance already exists in the United States.[110] Dialogue between countries can take place where cooperation between states is crucial, like in military technology and arms development, over areas where it may be more appropriate to respect a plurality of approaches. As in the case of the previous nuclear weapons ban treaty, overlapping consensus on norms and practical guidelines can also be achieved even when countries have different political or ethical considerations for justifying these principles. The delineation of global standards for AI ethics and governance should be informed by diverse cross-cultural perspectives that are attuned to the needs and desires of different populations.

## The Digital Divide and Decolonial AI

Not all technological growth is happening at the same place and pace, with the global digital divide deepening the gap between under-digitized and highly connected countries. According to an estimate from PwC, seventy percent of the 15.7 trillion-dollar wealth generated by the growth of AI will be shared between the US and China.[111] The two countries currently own more than 75% of the cloud computing market and hold close to 90% of the market capitalization of the 70 largest digital platform companies.[112] On a global scale, low- and middle-income countries are likely to bear the brunt of the negative impacts of AI, with little of the returns. Following a report by the United Nations Educational, Scientific and Cultural Organization (UNESCO), low and middle-income countries (LMICs), such as Africa, Latin America, the Caribbean, Central Asia, and Small Island states, are underrepresented in the AI ethics debate, which has primarily been dominated by European, East Asian and North American countries.[113] As argued by Ushnish Sengupta, American and European companies largely dictate the culture of algorithm development.[114] Omitting LMICs in the conversation on AI ethics not only goes at the risk of neglecting local knowledge, cultural and ethical pluralism, it is also ignoring their demands for global fairness. Part of the challenge with AI development is ensuring that we are not repeating old patterns of extraction and exploitation, and instead of engaging in forms of equal collaboration between states.

Foreign technological investments in LMICs, wrapped under the banner of technical solutions for the developing world, are not necessarily beneficial catalysts for change. Technological development and data-driven economic initiatives in ways that mirror international development projects can serve to strengthen the current asymmetrical power dynamics in the global economy. Rumman Chowdhury has warned against the rising forms of algorithmic colonialism.[115] China's US $1 trillion belt and road initiative, which seeks to create new economic corridors, has led to the aggressive development and implementation of digital infrastructure in Africa. Digital technologies reproduce similar power dynamics to single-resource extraction during the colonial area, whereby the harvesting, mining, extraction, measuring, and storage of people's data all lie in the hands of foreign investors. These new technologies also serve to strengthen the power and control of local governments, who want to attract these foreign investments. The Chinese tech company CloudWalk, with the promise of creating 'Safe Cities,' has implemented CCTV cameras with facial recognition technology in cities across Africa, leaving residents with no ownership over their faces or movements.[116] For Chowdhury, this "surveillance ecosystem is characterized by private or foreign actors controlling the digital civic backbone."

Another case of algorithmic colonialism is the rapid growth in mobile to mobile currencies and digital lending apps in Kenya's Silicon Savannah. Mobile phone-based lending systems were once praised as enfranchising East Africans who previously could not access banking and micro-finance. As stated by Birhane, since the 1990s, fintech is lauded as the technological revolution that will lift Africa out of poverty.[117] The ballooning industry has allowed poor Kenyans to get quick access to short-term loans, but many are now in a state of perpetual debt. As Chowdhury points out, these infrastructures of indebtedness have replaced the old kinship networks through which people borrowed and lent funds. The former peer-to-peer and reciprocal credit systems allowed for flexibility in lending, where margins and timelines could be renegotiated, and your credit score was assessed through trustworthiness in social relationships. Lending apps have eradicated these past social transactions, and lenders now rely on assessments from an impersonal and foreign algorithm. Lending apps have created what researchers Kevin P. Donovan and Emma Park termed "a novel, digitized form of slow violence that operates not so much through negotiated social relations, nor the threat of state enforcement, as through the accumulation of data, the commodification of reputation, and the instrumentalization of sociality."[118]

Much like old colonial patterns, much of the wealth generated by these "revolutionary" technologies goes back into the pockets of North American, European, and now Chinese companies. Even if Safaricom is 35% owned by the Kenyan government, Birhane points out that 40% of the shares are owned by UK multinational Vodafone and the other 25% by foreign investors.[119] Rather than lifting the nation's most impoverished, the emergence of FinTech is profiting from their state of indebtedness. As summarized by Birhane, "not only is Western developed AI unfit for African problems, the West's algorithmic invasion simultaneously impoverishes development of local products while also leaving the continent dependent on Western software and infrastructure." Many of these foreign investment projects operate in ethically questionable ways. As documented by Amy Hawkins, the Zimbabwean government agreed, without any public consultations, to send data on millions of black faces to CloudWalk Technologies to help train their systems towards darker skin tones.[120] Foreign investments in LMICs, driven by profit maximization, can take advantage of countries with laxer technology laws to exploit these "human natural resources." AI can perpetuate and legitimize global and local inequalities to a scale and scope no technology has ever done before.

Concerns over repeating old processes of colonization with new technologies, such as the exploitation and disenfranchisement of local populations, especially arise in the field of biotech. For Mark Maguire, the rise of biometric security is an eerily reminder of the use of fingerprinting in colonial contexts, to track and monitor local populations.[121] While governments have promised that biometric identity programs, such as the Aadhaar identity card in India, can allow impoverished communities to access government services or travel across borders, many are fearful of the potential for greater state surveillance and oppression. The Aadhaar card has also been accused of pushing poor Indians into starvation, as their basic food rations were not adequately linked to their identity cards.[122] Maguire notes how the modern field of biometrics, and facial recognition technology, still relies on 20th-century configurations of race. Variations in skin pigmentation and face geometry, though measurable and quantifiable, are grounded in sociological understandings rather than science. The racialization of a population is inevitably tied to the securitization of individual identities. In South Africa, a country long divided by the apartheid system, CCTV surveillance cameras have been accused of repeating old historical patterns that sought to exclude and segregate disadvantaged black communities from certain areas and spaces. Vumacam, an AI surveillance company, has been accused of profiling individuals in a manner akin to the apartheid era passbooks, with their software repeatedly flagging black construction workers for "suspicious behaviour."[123]

There are countless examples where colonial practices of oppression, exploitation, and dispossession are present in AI systems. For one, there are the unethical working conditions of ghost workers in the developing world, who are given data labelling and annotation tasks. The beta-testing and fine-tuning of AI systems are also part of a phenomenon known as "ethics dumping," whereby AI developers purposely test and deploy their technologies in countries with weaker data protection laws or onto vulnerable populations.[124] Cambridge Analytica infamously tested its voter manipulation systems in Kenya during both its 2013 and 2017 elections.[125] The New Zealand government first deployed its child welfare intervention algorithms on the Maori, an indigenous nation, which has long been the target of racism.[126] Indigenous people are still excluded as participants in the development of AI systems, despite the critical ramifications these systems have on their wellbeing and livelihoods. While an exhaustive list cannot be provided here, AI is clearly part of the norm, rather than the exception, when it comes to global inequities.

AI technologies are inextricably tied to the patterns of power that characterize our political, economic, and social worlds. The power dynamics between the world's advantaged and disadvantaged, instilled during the colonial era, are present in the design, development, and use of AI technologies. To protect and prevent harm against vulnerable groups, Mohamed, Png & Isaac[127] argued for a decolonial critical approach in AI to gain better foresight and ethical judgment towards advances in the field. Decolonial theory can unmask the values, cultures, and powers dynamics at play between stakeholders and AI technologies. Critical science and decolonial theory, when used in combination, can pinpoint the limitations of an AI system and its potential ethical and social ramifications, becoming a "sociotechnical foresight tool" for the development of ethical AI. As AI will have far-reaching impacts across the social strata, a diversity of intellectual perspectives must be a part of its development and application. Those who have the most to lose from AI systems are often readily aware of the social inequities they live and face.

## Capturing the Human - Data Collection & Ownership

With technologies of mass surveillance, there is an assumption that all communities and individuals have fallen under an AI magnifying glass. Researcher and visual artist Mimi Onouha illustrated how, despite the vast development and deployments of instruments of surveillance, measurement, and quantification, many data sets remain missing.[128] Certain areas of the world continue to evade Google maps, one example being Rio favelas who, although they are visible in the satellite view, disappear in map view. Only 26 of the city's 1000 favelas are mapped yet are home to around a quarter of the city's population.[129] By being off the grid, these populations are cut off from services, like mail delivery, garbage collection, electricity. It also has implications for the development of other technologies. Uber or other transportation apps do not service these areas. What3words is one among the geolocation apps looking to change this dynamic and render previously unmapped areas or non-identified zones legible and accessible. The apps intend to replace postcodes, addresses and difficult-to-follow directions by identifying each of the world's 57 trillion 3mx3m squares with a distinct combination of three words.[130] What3words is used by UK emergency services to rescue people in remote areas. Despite these positive developments, it is not only places but people that are also missing. Onouha has created a list of missing data sets[131], which ranges from poverty and employment statistics that include people who are behind bars to trans people killed or injured in instances of hate crime.

For Onouha, "measurements extracted from the flux of the real '', meaning that data sets are only a snapshot in time of a continually moving reality. When we obtain the data, we tend to forget the process and the relationships that went into creating it. We extract data in ways that fit our already existing patterns of collection. Patterns of absence are created when the data is hard to collect or where there is a lack of incentives, such as incidents of police brutality. Sometimes, avoiding data collection is a strategy as those who are missing in data sets often know that they are. One clear example is undocumented migrants living in the USA. Even individuals who find themselves unwillingly under the magnifying can find creative and ingenious ways to resist. New technologies can also spark acts of defiance. In the Occupied Palestine Territories, the extensive Israeli surveillance system, aimed mainly at tracing Palestinian dissidents, also allowed activists to post and share the faces of Israeli soldiers to demand greater public accountability in military operations.[132] In Saudi Arabia, the notifications from the app used to track women's movements under the male guardianship system were shared by feminists on Twitter to denounce the oppressive technology.[133]

Given the large volumes of quantification and the imbalances in data collection and consent, social scientists, such as Tahu Kukutai and John Taylor, have turned their attention to the notion of data ownership.[134] They present how First Nations have claimed their right to data sovereignty, which is the idea that data is subject to the laws and governance structures of each nation where it is collected. Many Indigenous groups have demanded their right to govern the collection, ownership, and application of their data following the principles stated in international instruments like the UN Declaration on the Rights of Indigenous Peoples.[135] Data ownership not only entails respect for Indigenous forms of knowledge but allows for their data to be used in culturally appropriate and culturally sensitive ways. As underlined by Maggie Walter, indigenous statics in Australia provided by the state are heavily characterized by terms such as disparity, deprivation, disadvantage, dysfunction, and difference, with little indigenous-owned data available to counterbalance this narrative.[136] Better Indigenous data collection and ownership will not only make Indigenous people more visible but also help to dismantle statistically informed pejorative stereotypes. This reflection should extend to other historically marginalized communities that do not fit within the dominant population model of the state[137].

## Rethinking AI in Labour & Medicine

There are other fields where ethnographic insights will be valuable in shaping AI ethics, two among them being labour and medicine. With the rise of automation and AI-powered systems, we appear to be ushering into a technological revolution where the future of human labour is increasingly insecure. Economists have made claims that more than 47% of American jobs will be rendered obsolete in 2030.[138] Digital technologies, such as smartphones, have blurred the boundaries and responsibilities of employees and employers. Online career networks such as LinkedIn have reframed the ways we present ourselves as workers; we fashion our appearances and claims to seem hirable. The rise of the gig economy and digital nomads have opened possibilities for carrying out work from anywhere in the world, but also new forms of uncompensated labour. Scholars like Mary Gary and Siddharth Suri have documented the hidden human content moderators whose task is to keep obscene content out of users' feeds and the continuation of ill-protected and unrecognized labour in the automation revolution.[139] An anthropology of labour reminds us that algorithmic operations are often more human than they appear, and of the importance of examining the people working beside and within these systems.

Culture will play a factor in determining how we welcome these technologies, in which individuals must participate in the labour economy and what type of tasks are valued forms of labour. Scholars like Braden Molhoek predict that advancements in AI will allow humans to transfer "hazardous and undesirable jobs" to intelligent machines, hence allowing individuals in society to "pursue other activities or pleasures."[140] A study by Schliewe and Petzoldt on self-scan checkouts showed how Russian participants had higher levels of social pressure in adopting these technologies despite higher rates of technological anxiety and lower level of self-efficacy.[141] In contrast, German participants did not feel any social pressure or a greater inclination to use these instruments despite their higher efficacy. Automation efforts, such as self-scan checkouts, often serve as cost-shifting or cost-saving measures, rather than ridding humans of hazardous or dull labour. For researchers Ming-Hui Huang, Roland Rust, and Vojislav Maksimovic, the expansion of AI can change the types of labour we value within society.[142] They predict the rise of a "feeling economy," where AI performs the analytical and thinking tasks, and human workers are responsible for interpersonal and empathetic tasks (e.g., communication, maintaining relationships, influencing others), placing more emphasis on the emotional dimensions of their work.

Enthusiasm over the transformation in labour brought on by the development of AI varies from one country to another. In societies where strong social safety nets are in place, such as Sweden, workers tend to express more positive attitudes towards job automation.[143] In the US, the trend is the reverse, where there is greater anxiety over job loss and economic inequality.[144] While we have witnessed a growth in freelancing careers and the gig economy, workers fear they will have to take on multiple jobs to afford to live in a future tech-dominated economy. There is also the alarming problem that contract workers and the self-employed often fall through the cracks of social protection programs and unemployment benefits. Economists further predict a widening of the gaps between low and high skilled workers, with a more prominent and higher-paid managerial class. Greater global interconnectedness has allowed for new types of contract workers to emerge. An early case study on transnational labour by Nardi and Kow examined how Chinese workers that perfect computer games would collect coins to sell to other less skilled game players abroad, a phenomenon they titled "gold farming."[145]

Rather than causing human labour to evaporate seemingly, AI systems will continue to require substantial amounts of human labour to be successfully integrated and new skills from those who work alongside them. Shestakofsky, through his ethnographic observations at tech-startup AllDone, demonstrates how human workers continue to work alongside automated machines.[146] At the first stage of its development, when AllDone was looking to attract users to its system, it relied upon the efforts of a Filipino contract team to collect information and to target potential users and to conduct a digital marketing campaign. This team of workers provided what termed Shestakofsky "computational labour" when software engineers at AllDone did not have the resources to design and develop an automated system. After acquiring a customer base, AllDone turned its attention to securing sellers on the website. Many of the sellers were small entrepreneurs or individual workers that did not understand the design and rules of the system and often voiced frustration over a lack of responses to quotes. To help repair this knowledge gap, AllDone hired a team of customer service agents that would patiently explain the system to new sellers and offer them advice on how to improve their profile. This team provided the emotional labour needed to help users adapt to the system. In its third stage, when AllDone sought to extract higher profits from its users and sellers, it relied on both emotional labour to convince users to keep their subscription and computational labour to prevent sellers from circumventing and gaming the new rules.

Even with new technologies and expectations of changes, longstanding trends, and divides in labour continue to persist. The emotional labour of counselling potential sellers and reassuring customers was accomplished by the women who led the phone line support staff. Women have long been characterized as support and care workers. These perceptions are replicated in technologies and within the tech industry. Cheap and disposable labour is extracted from low and middle-income countries, where underpaid Filipino workers are given short-term contracts to fill in when computational resources are missing. The expansion of an informal AI labour sector, with poorly paid click workers that trace and label the photos of machine learning, is growing across the Global South.[147] Their efforts, in comparison with North American working and wage standards, are mostly uncompensated. Outsourcing, offshoring, and contracting labour to the Global South will continue to persist even in a world of software automation, with underpaid workers providing human assistance needed to create and keep running these systems and the managerial team accruing wealth. Rather than pushing humans out of production, technology-driven automation may be replicating and reinforcing inequities at the domestic and global scale.

There is a recurrent tendency to obscure the human labour needed to integrate these technologies, with their adoption necessitating new skills, new routines, and changes to the physical infrastructure.  In the booming agrotech sector in the US, farmers were pushed to reconfigure the physical infrastructure of their barns and silos so that they would be more apt to sensor readings and data collection.[148] US farmers often felt that agrotech developers had no first-hand knowledge of agricultural practices. Seemingly mundane issues like securing rural broadband internet and learning to use and interpret the data behind these technologies was a significant barrier to farmers[149]. Due to the cyclical and weather-dependent nature of farming, it was difficult for farmers to correlate investments in these new technologies with higher crop yield. The integration and success of these agrotech tools were dependent on the social conditions, financial resources, and labour-power of farmers. For smaller family farms, their inability to undertake the financial risks of adopting these systems left them at a disadvantage against large finance-backed conglomerates who could easily absorb the costs of experimentation. Larger agrotech companies, financial institutions, and agricultural conglomerates are the ones who will profit from these technologies, leaving small farmers behind and vulnerable to exploitation by these new tech intermediaries and vendors.

Another field in which artificial intelligence is predicted to have a significant impact is in medicine. The development of health tech promises to reduce several of the errors that have led to aggravated patient risks and provide better and more accurate medical diagnoses, such as the early detection of abnormal cell growths. Medical AI has also been cited as having the potential to improve hospital workflows and allow patients to process their health data.[150] The field of medical anthropology quickly started to assess the impacts of health technology on patient wellbeing and hospital environments. Rapid ethnography is now more commonly applied in evaluating the design and implementation of user-oriented health IT devices and systems. Anthropological studies in health technology have examined how such devices are changing relationships between health care workers and patients, and how social relations between healthcare workers are changed.

One of the underlying threads in the application of health technologies is how they fail to consider the environment in which they are integrated. Researchers Ackerman, Gleason & Gonzalez have demonstrated how IT designers can build quality devices but lack an understanding and engagement with the workplaces and ethics in which they are integrated.[151] They use the example of the failure of a free-standing urinary tract infection (UTI) kiosk, whose goal was to reduce patients' wait times for clinical services in urgent-care clinics. In trial tests, the kiosk was proven to be efficient and high patient satisfaction. When the technology was integrated into emergency departments, they proved to be a total failure due to staff resistance and lower than anticipated patient eligibility for kiosk-assisted care. On top of the issue was finding the right spot for the kiosk. Nurses who were in charge of patient triage in hospitals felt that the kiosk was a time-consuming detour and, most importantly, it would signal outpatients who they believed to have contracted a UTI, raising issues of patient and medical confidentiality. For Ackerman et al., such studies are useful in demonstrating how "the interests of tools and users are not always well-aligned." Indeed, many healthcare workers are suffering from burnouts due to the prevalence of electronic health records and other technologies that are aimed at making the healthcare industry more data-driven.[152]

Although the refusal to use a self-standing UTI device may appear as a relatively benign example, other ill-fitting AI-driven medical diagnosis systems can have grave consequences on patient wellbeing. Eran Tal offers the case of fibromyalgia, a chronic pain condition for which no clinically established biomarkers exist and whose diagnosis procedure largely rests on a patient's testimony. Computer scientists are trying to develop an AI system to diagnose fibromyalgia by analyzing data on individual neural signatures.[153] For Tal, "Machine learning algorithms may eventually become 'gold standards' for diagnosing some diseases, like fibromyalgia, that is difficult to diagnose in traditional ways."[154] Giving the final say to the algorithm can gravely harm individuals who were previously eligible for treatment by artificially restricting the category of the disease. Medical anthropologists are starting to reflect on what regulatory structures ought to be in place when using the results produced by AI to ensure patient interest and wellbeing. American cardiologist Eric Topol argues that, paradoxically, machine-led and digital medicine can help make healthcare more human. He states that AI can "giving doctors and patients the gift of time – to get back to where medicine was decades ago when the relationship was characterized by a deep bond with trust and empathy."[155] He nonetheless cautions that the rise of medical AI, if it continues to be driven by private and commercial interests, risks deteriorating doctor-patient relationships and exacerbating the working conditions of medical practitioners.[156]

## Conclusion

Following Hagerty and Rubinov, we must remember that "AI systems are developed and used in an imperfect and unequal world, a place where disastrous things have taken place and could just so easily happen again."[157] Discrimination, dispossession, and exploitation by AI systems are already occurring in both the domestic and global landscape. To avoid harming the world's disadvantaged and further emboldening its privileged, computer scientists, social scientists, governments, and policymakers must work together on creating a globally beneficial future for AI. Our exploration of issues of privacy, fairness, user model manipulation offered a window into some of the challenges facing computer scientists, yet better algorithms can only go so far. Concentrating the conversation on models alone ignores the broader social, political, and cultural forces at play in the design, use, and implementation of these systems. As seen by the reproducibility crisis and the controversies facing Silicon Valley, those within the tech industry, whether consciously or unconsciously, are bound to certain academic, economic, cultural, and social patterns. These overlooked or unremarked patterns can cause substantial social disruptions and reinforce economic disparities on a local and global scale.

Anthropology is a discipline that embraces contradictions and recognizes that in almost every instance, new digital technologies can lead to benign and malign consequences. Through examining how tech is affecting culture and culture is transforming tech, we can start to get a better idea of the messy reality in which AI systems are being created and deployed. AI has the power to automate the most undesirable aspects of our world. This technological revolution offers a unique opportunity to reassess what are our human values, what it means to live together in society, and how to have different voices heard. Anthropology encourages us to question our most taken-for-granted cultural and social assumptions about how our world operates. With its inherently critical mindset, the discipline can help predict the areas in which AI systems are likely to wreak havoc and go back to the humans shaping these developments. It can contribute to the development of ethically plural and culturally sensitive technologies, and in turn, AI principles and regulations that can break longstanding global inequities.

Anthropology can also help evaluate and assess the impact of AI in other fields, such as international development, labour, and medicine. Against the techno-dystopian view that AI will spell the end of humanity, it can help remind us of what it means to be human in the first place.

[26] (n.d.). The Algorithmic Foundations of Differential Privacy - CIS UPenn. Retrieved August 29, 2020, from https://www.cis.upenn.edu/~aaroth/Papers/privacybook.pdf

[27] Cormode, G., Jha, S., Kulkarni, T., Li, N., Srivastava, D., & Wang, T. (2018, May). Privacy at scale: Local differential privacy in practice. In Proceedings of the 2018 International Conference on Management of Data (pp. 1655-1658).

[28] Kearns, M., & Roth, A. (2020). *The ethical algorithm: The science of socially aware algorithm design*. New York, NY: Oxford University Press.

[29] https://www.wired.com/story/strava-heat-map-military-bases-fitness-trackers-privacy/

[30] (n.d.). 2019 AI Predictions: Six priorities you can't afford to ignore: PwC. Retrieved August 29, 2020, from https://www.pwc.com/us/en/services/consulting/library/artificial-intelligence-predictions-2019.html

[31] Garg, N., Schiebinger, L., Jurafsky, D., & Zou, J. (2018). Word embeddings quantify 100 years of gender and ethnic stereotypes. Proceedings of the National Academy of Sciences, 115(16), E3635-E3644.

[32] Kearns, M., & Roth, A. (2020). *The ethical algorithm: The science of socially aware algorithm design*. New York, NY: Oxford University Press.

[33] https://www.washingtonpost.com/business/2019/11/11/apple-card-algorithm-sparks-gender-bias-allegations-against-goldman-sachs/

[34] Kleinberg, J. (2018, June). Inherent trade-offs in algorithmic fairness. In Abstracts of the 2018 ACM International Conference on Measurement and Modeling of Computer Systems (pp. 40-40).

[35] (n.d.). Pareto Front - Cenaero. Retrieved August 29, 2020, from http://www.cenaero.be/Page.asp?docid=27103&langue=EN

[36] Kearns, M., Neel, S., Roth, A., & Wu, Z. S. (2018, July). Preventing fairness gerrymandering: Auditing and learning for subgroup fairness. In International Conference on Machine Learning (pp. 2564-2572).

[37] Foulds. J. R. , Islam R., Keya K.M., & Pan S.(2019, September 10). An Intersectional Definition of Fairness. Retrieved August 29, 2020, from https://arxiv.org/abs/1807.08362

[38] Ibid.35